\documentclass[usegraphicx,useAMS,usenatbib]{mn2e}
\usepackage{graphicx}

\title[Radii of M-dwarfs]
 {The radii of M-dwarfs in the young open cluster NGC~2516}

\author[R.J. Jackson et al.]
  {R.J.~Jackson, R.D.~Jeffries and P.F.L.~Maxted\\
  Astrophysics Group, Research Institute for the Environment, Physical
  Sciences and Applied Mathematics, Keele University, \\ Keele, 
      Staffordshire ST5 5BG
}
\date{Submitted July 1 2009}

\pagerange{\pageref{firstpage}--\pageref{lastpage}} \pubyear{2009}

\def\LaTeX{L\kern-.36em\raise.3ex\hbox{a}\kern-.15em
    T\kern-.1667em\lower.7ex\hbox{E}\kern-.125emX}

\begin{document}
\label{firstpage}
\maketitle

\begin{abstract}
Using a novel technique, which combines previously determined rotation
periods with new spectroscopic determinations of projected rotation
velocity, we have determined radii for fast rotating, low-mass
(0.2--0.7\,$M_{\odot}$) M-dwarfs in the young, solar-metallicity open
cluster, NGC~2516.  The mean radii are larger than model predictions at
a given absolute $I$ or $K$ magnitude and also larger than the measured
radii of magnetically inactive M-dwarfs; the difference increases from
a few per cent, to 50 per cent for the lowest luminosity stars in our
sample. We show that a simple two-temperature starspot model is broadly
capable of explaining these observations, but requires spot coverage
fractions of about 50 per cent in rapidly rotating M-dwarfs.
\end{abstract}


\section{Introduction}
The radius of a star is among the most fundamental properties that can
be predicted by a stellar model. Yet recent measurements of radii
for K- and M-dwarfs in eclipsing binary systems reveal
significant discrepancies between models and observations. For a given
mass, the radii of stars with $0.3<M/M_{\odot}<0.8$ are up to 20 per
cent larger than predicted by evolutionary models, and so the effective
temperature at a given luminosity is over-predicted by up to 10 per
cent (L\'opez-Morales 2007; Morales et al. 2009).  Attempts to explain
these discrepancies have focused on the role of magnetic fields in
suppressing convection and producing cool starspots (D'Antona et
al. 2000; Mullan et al. 2001; Chabrier, Gallardo \& Baraffe 2007).
Strong, dynamo-generated magnetic fields are expected in the tidally
locked, rapidly rotating components of eclipsing binaries, but if these
effects are important they should also cause larger radii
for active, single stars.  Confirming this is challenging because (i)
the small number of low-mass stars close enough to measure precise
interferometric radii are magnetically in-active, and (ii) metallicity
variations, which are difficult to accurately assess in cool field
dwarfs, could be a significant source of radius scatter (Berger et
al. 2006).

In this letter we present radius estimates for rapidly rotating,
and hence magnetically active, single, cool dwarfs in the young, open
cluster NGC~2516, which has a close-to-solar metallicity. Our novel,
purely geometric technique uses spectroscopic measurements of projected
equatorial velocities along with rotation periods
from light curve modulation to infer projected stellar radii.  
Our targets have masses $0.2<M/M_{\odot}<0.7$, covering spectral
types M0--M5, reaching masses at which stars are expected to
be fully convective. We show that, like the low-mass components of
eclipsing binaries, these stars have significantly larger radii than
model predictions and that widespread coverage with dark, magnetic
starspots is a plausible explanation.

\section{Observations of NGC~2516}
\label{observations}

NGC~2516 is a young open cluster with a large population of low-mass
stars (Hawley, Tourtellot \& Reid 1999; Jeffries, Thurston \& Hambly
2001). It has a close-to-solar metallicity, spectroscopically
determined as [Fe/H]$=0.01\pm 0.07$ or photometrically as [M/H]$=
-0.05\pm 0.14$ (Terndrup et al. 2002). The same authors give an intrinsic distance
modulus of $7.93\pm 0.14$ based on main sequence fitting.
The cluster age is $\simeq 150$\,Myr based on the nuclear
turn-off in its high-mass stars and the lithium depletion and X-ray
activity seen in cooler stars (Jeffries, James \& Thurston 1998; Lyra
et al. 2006). The cluster reddening is $E(B-V) = 0.12 \pm 0.02 $
(Terndrup et al. 2002).

Targets in NGC~2516 were observed using the VLT (UT-2 Kueyen) FLAMES instrument
feeding the GIRAFFE and UVES fibre spectrographs. Primary targets
were chosen from 362 candidate cluster members reported by Irwin et
al. (2007), that have measured rotation periods in the range
0.1-15~days and $14.5<I<18.5$, corresponding approximately to
$0.2<M/M_{\odot}<0.7$ (Baraffe et al. 2002). Spare fibres were placed
on ``blank'' sky positions and on other photometric candidates for
which no rotation periods were known. The GIRAFFE spectrograph was used
with the HR20A grating, covering the wavelength range 8000--8600\AA\ at
a resolving power of 16\,000.  Bright blue stars in the same field were
used for telluric compensation and observed simultaneously using UVES. 
Eight observation blocks (OBs) were recorded in service mode between 27
November and 30 December 2007. Each OB consisted of two 1280\,s
exposures with GIRAFFE and three 800\,s UVES exposures. One of the OBs
was performed twice sequentially with no change in instrument
setup. Some targets were also observed in multiple
OBs. A total of 867 spectra were recorded of 678 unique targets.

\subsection{Extraction of the target spectra}

Many of the spectra were faint, requiring optimal extraction
to provide sufficient signal-to-noise ratio (SNR) for useful analysis.  There was
also significant telluric absorption and strong sky emission lines in the
spectra. For these reasons we used purpose-built software for data reduction
rather than pipelines supplied by the European Southern Observatory.
 
The method used for optimal extraction is described by Horne
(1986). This applies non-uniform weight to pixels in the extraction
sum, minimising statistical noise whilst preserving photometric
accuracy. In our case, weighting profiles were determined from a boxcar average
along the wavelength axis of tungsten-lamp flat field images.
Wavelength calibration was based on thorium-argon lamp spectra recorded
during the day.  Fine offsets were applied to the calibration of
individual fibres to compensate for any drift with time. These offsets
were determined by comparing the positions of emission lines in the
median sky spectra of each OB to their position in sky spectra averaged
over all OBs. The offset per fibre varied between 0.001\AA\ and
0.008\AA.  To deal with variations in fibre efficiency, the proportion
of the median sky spectrum subtracted from each target spectrum was
tuned to minimise the peak amplitude of the cross-correlation between
the sky-subtracted target spectrum and the median sky spectrum. Spectra
from each OB were averaged and corrected for telluric absorption using
the co-temporal UVES spectra of bright blue stars, which were broadened to
mimic the GIRAFFE spectral resolution.

\subsection{Measurement of radial and projected equatorial velocities}

Heliocentric radial velocities (RVs) were measured by cross-correlation
with stars from the UVES atlas (Bagnulo et al. 2003).  Templates of
type K4.5V (HD~209100) and M6V (HD~34055) were used. Projected rotation
velocities ($v \sin i$, where $i$ is an unknown inclination) were
estimated from the cross-correlation function widths. Translation from
these to $v\sin i$ was calibrated by artificially broadening standard
star spectra. This was done in two stages. The standards were broadened
with a Gaussian to match the cross-correlation function widths to the
average width obtained from 40 slow-rotating targets (period, $P>5$
days), which we expect to have negligible $v \sin i$ (compared with the
spectral resolution).  These broadened spectra were then convolved with
a rotational broadening kernel at many $v \sin i$ values.  A linear
limb darkening coefficient of 0.6 was used (Claret, Diaz-Cordoves \&
Gimenez 1995), but our results are quite insensitive to this parameter.
The $v \sin i$ values found using the two templates did not differ
significantly, either on average or as a function of colour, so we adopted the
average value.

Uncertainties in RV and the width of the cross-correlation function
were measured empirically. Uncertainties due to noise in the target
spectra were estimated, as a function of SNR, by comparing results for
the two sequential OBs.  A constant representing other
error sources was determined by comparing observations of the same
targets through different fibres in different OBs. Typical RV
uncertainties were 0.2--0.6\,km\,s$^{-1}$ and the
normalised uncertainty in $v \sin i$ was typically $\simeq 10$ per cent for
$v \sin i >20$\,km\,s$^{-1}$ but larger for smaller $v \sin i$.

\begin{figure}
	\centering
		\includegraphics[width=0.42\textwidth]{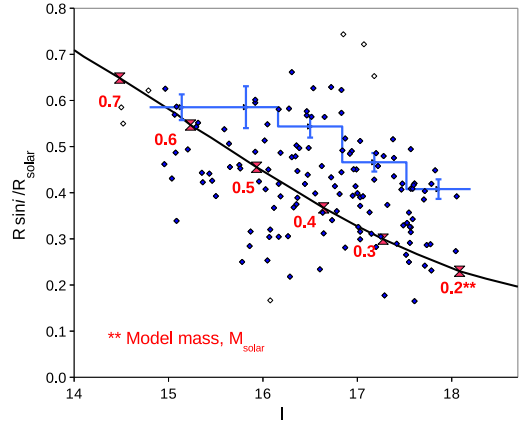}
	\label{fig:NGC2516fig1}
		\caption{Measured radii of stars in NGC~2516 as a function
		of $I$-magnitude. Diamonds show the projected radii
		($R\sin i$), solid diamonds were
		used to determine the average radii. The histogram
		shows the average radius determined from the mean
		$R\sin i$ in each bin. The solid curve is a 158\,Myr, solar
		metallicity isochrone from the Baraffe et al. (2002) models.}
\end{figure}

\section{Average radii of cluster members}

The NGC~2516 photometric candidates used for the radius estimates 
have $v \sin i>8$\,km\,s$^{-1}$, and most have periods $<3$\,days (see below). 
This sample is unlikely to contain many field stars, because fast rotators are rare
among the general late-K to mid-M-dwarf field population (Delfosse et al. 1998).
Nevertheless, some candidates might be tidally locked binaries seen at
an unfortunate phase or have low quality measurements of $v\sin
i$.  To avoid contamination of the sample we excluded: stars with
spectra of average SNR less than 5; targets with cross-correlation
functions with no clear single peak, or multiple peaks characteristic
of a binary pair; stars with measured $v \sin i$ less than
8\,km\,s$^{-1}$ or less than twice their uncertainty. Stars were also excluded with RV
more than $2(\sigma_{v}^{2} + \sigma_{c}^{2})^{1/2}$ from the mean ensemble
 RV (for all the included stars with measured $v \sin i$), where
$\sigma_v$ is the individual RV uncertainty and $\sigma_c$ is the
intrinsic cluster RV dispersion, determined to be $\sigma_c = 0.66\pm
0.13$\,km\,s$^{-1}$. The latter step was performed iteratively.

The remaining 147 targets were used to determine projected
radii using the formula $R \sin i/R_{\odot} = 0.124\, (P/2\pi)\, v\sin
i$, where $P$ is in days and $v \sin i$ is in km\,s$^{-1}$.
Figure 1 shows projected radii as a function of $I$-magnitude. The
photometry comes from Irwin et al. (2007), with a a small correction
($\Delta I = 0.080-0.0076\, I$\,mag) added, to put Irwin et al.'s
magnitudes onto the better-calibrated photometric scale established by
Jeffries et al. (2001) for low-mass stars in NGC~2516.

Assuming stellar spin axes are randomly orientated, then 
the mean value of $R\sin i$ measured over a small range of
$I$-magnitudes can be divided by the average value of $\sin i$ to
determine the mean radius. A complication is that it 
is not possible to measure the
period of stars, or determine $v \sin i$ values if they are viewed at
small inclination angles, resulting in an upward bias in the mean $\sin
i$ (see Jeffries 2007). If we consider a cut-off inclination, $\tau$,
such that stars with $\sin i<\tau$ yield no period or $R \sin i$ value,
then the normalised probability density of $i$ becomes
\begin{equation}
P(i) = \sin i / \cos(\arcsin \tau) \ \ {\rm for}\ \tau < \sin i < 1\, ,
\label{probdensity}
\end{equation}
and the mean value of $\sin i$ becomes
\begin{equation}
\int \sin i\, P(i)\, di = \tau/2 + (\pi/2 - \arcsin \tau)/2\cos
(\arcsin \tau)\, .
\label{meansini}
\end{equation}
Equation~\ref{meansini} can be used to determined the mean radius from
a set of $R \sin i$ values providing $\tau$ is known. Here $\tau$ is
estimated by forming a cumulative distribution of the measured values
of $R \sin i$ normalized with a regression line against $I$
magnitude. This is compared with the cumulative distribution predicted
(by Monte Carlo simulation) from the product of the modified $\sin i$
distribution of equation~\ref{probdensity} and a normal distribution
representing the measurement uncertainties in $R \sin i$, which is 
dominated by the error in $v \sin i$. A
Kolmogorov-Smirnov test is used to compare the measured and predicted
distributions for various $\tau$ values, yielding a best-fit $\tau =
0.40^{+0.06}_{-0.08}$, and hence an average value of $\sin i$ of
$0.832^{+0.015}_{-0.013}$ from equation~\ref{meansini}.  This compares
with a value of 0.785 if this effect were neglected, showing that it is
a small correction with an even smaller uncertainty when compared with the
size of the effects we will subsequently discuss.

\subsection{Variation of radius with I magnitude}
Figure 1 shows individual $R\sin i$ measurements and the average radius
estimated in five equally spaced bins for $14.8<I<18.2$. Uncertainties
in the binned mean radii are the quadrature sum of uncertainties in the
average $R\sin i$ and uncertainty in the mean $\sin i$. A few stars
were excluded from the binning process (but are still shown in Fig.~1),
because they either fell outside these magnitude ranges, or were more
than 2 standard deviations away from the regression line between $R
\sin i$ and $I$ discussed in the last subsection. Several of the latter
have long periods $>4$~days and may have incorrect periods at a 1-day
alias of the correct period (see Irwin et al. 2007), or over-estimated
$v \sin i$ because of unresolved binarity (see below).

Also shown in Fig.~1 is a solar metallicity 158\,Myr isochrone from the
Baraffe et al. (2002) models, which use a convective mixing length of 1
pressure scale height. The absolute $I$ magnitudes of the isochrones
were shifted according to a distance modulus ($7.93\pm 0.14$ --
Terndrup et al. 2002) and extinction ($A_I = 1.78 \,\, E(B-V) =
0.213$\,mag), with a net uncertainty of about $\pm 0.15$\,mag.  From
this diagram it appears that the cool stars observed in NGC~2516 have
larger radii (at a given $I$) than predicted by the models, by a factor
increasing from a few per cent at $I \simeq 15$ (corresponding to
$M \simeq 0.6\,M_{\odot}$), to about 50 per cent
at $I \simeq 18$ ($M \simeq 0.2\,M_{\odot}$). A plot can be
constructed for $R \sin i$ versus $K$ magnitude and a similar
discrepancy between observed and predicted radii is present.

\begin{figure}
	\centering
		\includegraphics[width=0.38\textwidth]{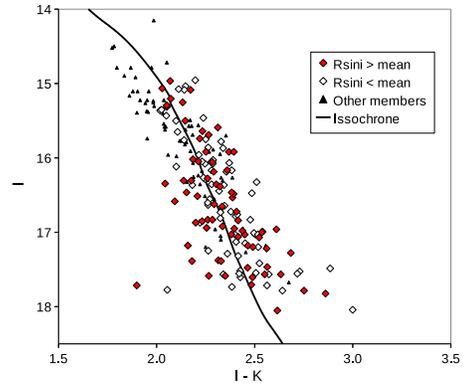}
	\caption{
A colour magnitude diagram for targets in NGC~2516. Solid diamonds show
	members with $R\sin i$ above the mean of their bin, open diamonds show
	$R\sin i$ below. Triangles are members with no $R \sin i$ measurement. 
	The line is a 158\,Myr solar metallicity isochrone (Baraffe et
	al. 2002).}
	\label{fig:NGC2516fig2}
\end{figure}

\subsection{Colour magnitude diagram}

\begin{figure*}
\centering
\begin{minipage}[t]{0.45\textwidth}
\includegraphics[width=75mm]{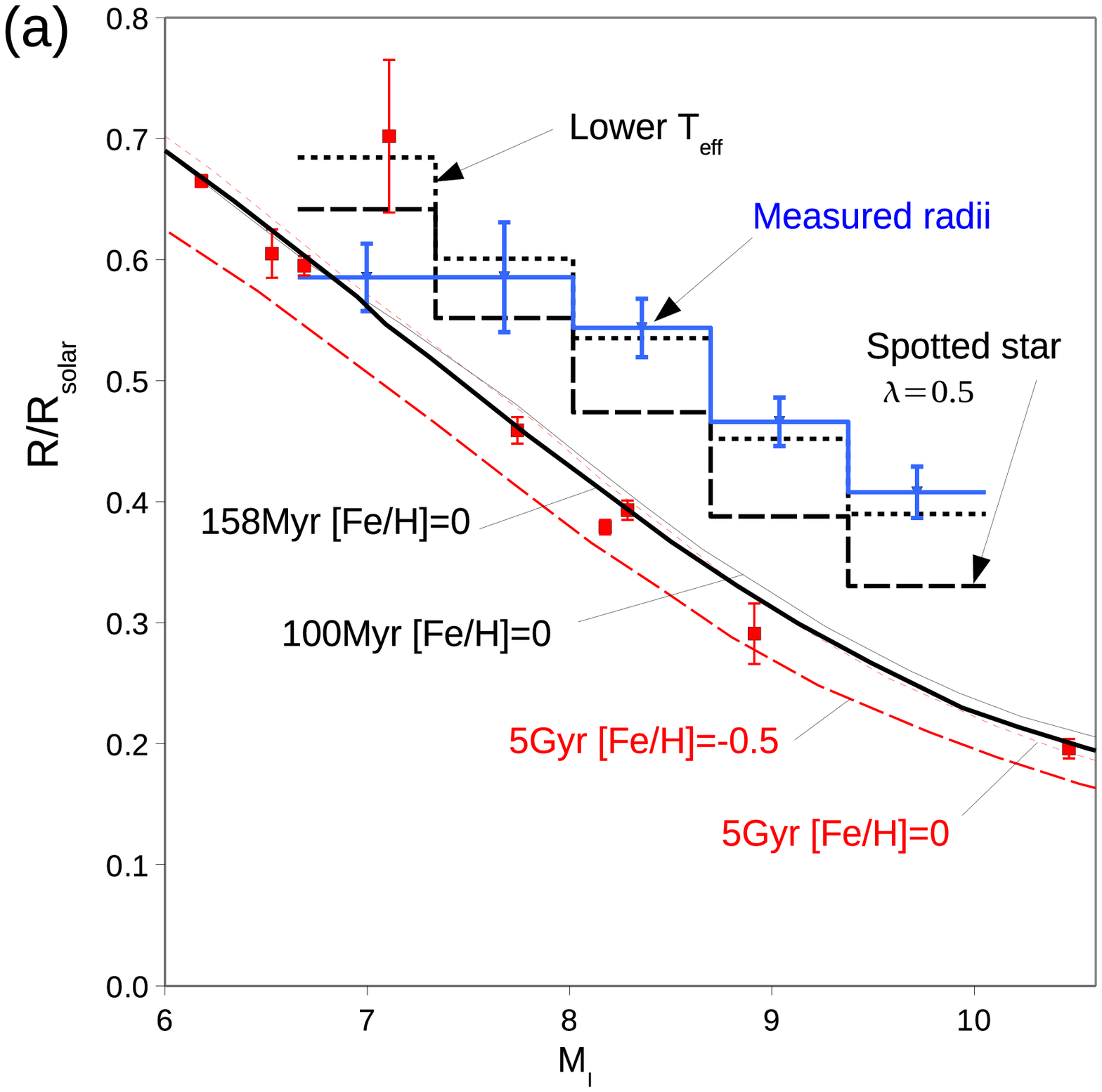}
\end{minipage}
\begin{minipage}[t]{0.45\textwidth}
\includegraphics[width=75mm]{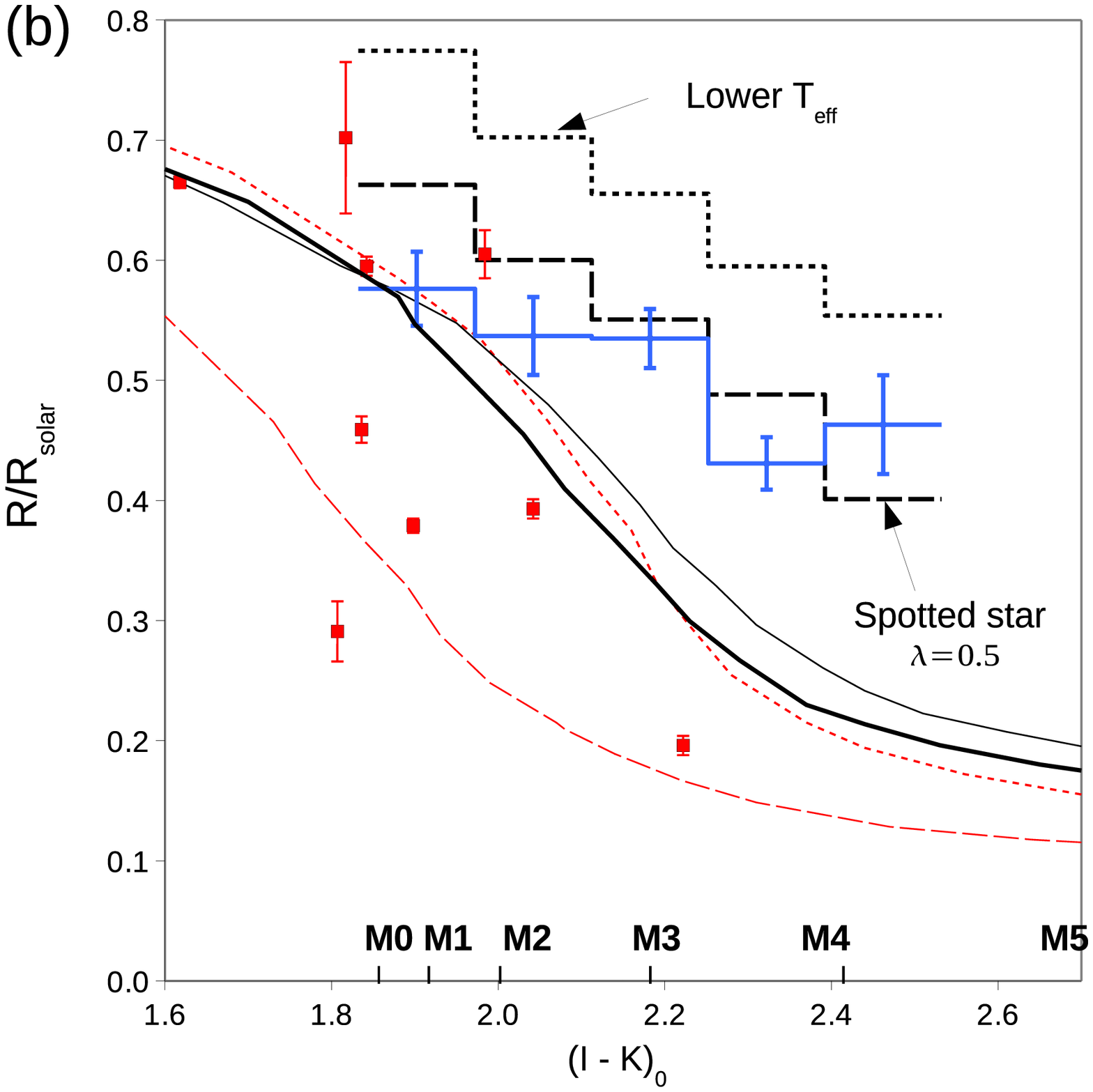}
\end{minipage}
\caption{Comparison of measured radii with model isochrones as a function of absolute $I$ and
intrinsic $I-K$. The solid histogram shows the mean value and expected uncertainty of measured
radii in NGC2516. Squares show directly measured radii for field stars. 
The labelled lines shows model isochrones for various ages and metallicities from Baraffe et al. 2002. 
The dotted and dashed histograms show the expected radius for (i) stars with a uniformly lower Teff 
and (ii)heavily spotted stars with 50 per cent spot coverage.}
\label{fig:NGC2516fig34}
\end{figure*}

The cluster may contain spectrally unresolved binary systems that are
seen at a phase/velocity-separation that results in them having the
cluster RV, but an over-estimated $v \sin i$ and hence over-estimated
$R \sin i$. Conversely, unresolved binaries will be shifted by up
to 0.75~mag leftward in Fig.~1, which would artificially reduce the
discrepancy between observed radii and models.

To check this, Fig.~2 shows the $I$ vs $I-K$ colour magnitude diagram
(CMD) for all target stars with SNR $>5$, where we identify stars with
$R\sin i$ above or below the mean of their bins in Fig.~1. $K$
magnitudes from the 2MASS catalog (Cutri et al. 2003) are transformed
to the CIT system using the correction ($K_{\rm CIT} = K_{\rm 2MASS} +
0.024$) given by Carpenter (2001). Compared to a 158\,Myr
isochrone from Baraffe et al. (2002), both classes of object lie
redward by an average of 0.06\,mag and are statistically
indistinguishable. Because we know that binaries where both components
contribute to the magnitude and observed spectrum should be shifted up
and to the right in this CMD, we can say that the
$R\sin i$ distribution appears unaffected by binarity.


\section{Discussion}

Figure~3 shows the mean radii for our NGC~2516 sample, binned as a
function of absolute $I$ and intrinsic $I-K$, compared with
various isochrones (from Baraffe et al. 2002, with a mixing length of 1
pressure scale height) and with data for single field stars.
Interferometric radii and parallaxes for these stars are from Demory et
al. (2009), $I$ magnitudes are from Leggett (1992) and $K$ magnitudes
are from 2MASS, converted to the CIT system.  The field stars have
low magnetic activity levels as judged from their coronal X-ray
emission, but a range of metallicities (L\'opez-Morales 2007). 
The 5\,Gyr isochrone for solar metallicity stars is in reasonable
agreement with the radii of inactive field stars (Demory et
al. 2009). Comparison of the measured and predicted radii as
a function of $I-K$ shows more scatter, but the majority of this
scatter can be interpreted as a metallicity
effect. The 5\,Gyr isochrone for [Fe/H]=-0.5 in Fig.~3b, shows that
metallicity strongly affects $I-K$, since, at a given radius, the $I$
magnitude is more sensitive to metallicity than $K$.

Turning now to NGC~2516, the metallicity complication is not present --
we know it has a close-to-solar metallicity (Terndrup et al. 2002).
The mean radii (as a function of $I$, $K$ or $I-K$) show an increasing
departure from the 158\,Myr solar metallicity isochrone with decreasing
luminosity and increasing colour (decreasing $T_{\rm eff}$). Even
reducing the assumed cluster age to 100\,Myr (a fairly firm lower
limit), does little to mitigate the discrepancy (see
Fig.~3). Uncertainties in the assumed extinction and reddening produce
the equivalent of only $\pm 3$ per cent systematic error in the
observed radii and the distance uncertainty is limited to shifting the
models in Fig.~3a by $\pm 0.14$\,mag along the x-axis.  We conclude
that the radii of low-mass stars in NGC~2516 do show departures from
model predictions and from the observed radii of in-active stars. The
radius discrepancies grow as we move to lower luminosities and mass,
becoming as much as 50 per cent larger at $M_I \simeq 10,\ M_K\simeq
7.5$, corresponding to $M \simeq 0.2\,M_{\odot}$ in the Baraffe et
al. (2002) models.

Direct comparison with other models of stellar evolution is more
difficult since these use empirical bolometric corrections to convert
$T_{\rm eff}$/luminosity to predicted colour/magnitudes. However, the
low-mass models of D'Antona \& Mazzitelli (1997) and Siess, Dufour \&
Forestini (2000) predict similar or even slightly smaller radii for a
given luminosity than Baraffe et al. (2002), so the discrepancy does
not just exist for this latter set of models.

\subsection{Effect of spot coverage}
The measured radii of the cool NGC~2516 stars are significantly higher
than predicted by evolutionary models. Similar discrepancies as large
as 20--30 per cent are reported for active stars in binary systems
(L\'opez-Morales 2007). Chabrier et al. (2007) discussed two possible
causes for this observed increase in radius. The first was a change in
mixing length in magnetically active stars affecting heat transfer. The
second was high levels of magnetic starspot coverage. They showed that
reductions in mixing length had only small effects for fully convective
stars where convection is nearly adiabatic. Here we find larger radius
anomalies for the cooler, fully convective stars, leading us to
consider surface spotting.

The NGC~2516 stars with measured $R \sin i$ have median rotation
periods of 0.76~days, with 95 per cent less than 3 days.  In either
single or binary M-dwarfs, this rotation should lead to saturated
levels of magnetic activity (e.g. James et al. 2000).
Cross-correlating our sample with the NGC~2516 targets in Hawley et
al. (1999), we find 27 matches. Of these, 11 stars have measured periods 
and $R\sin i$ values. All 11 show H$\alpha$ in emission 
effectively defining the upper envelope of magnetic activity in the cluster.

To quantify spot
coverage effects, a simple model assuming a fixed luminosity is used to
predict changes in magnitude and colour with spot temperature and area.
We start from the assumption that the Baraffe et al. (2002) models do a
reasonable job in predicting the radii, temperature and $I-K$ of inactive and presumably
unspotted stars (see Fig.~3 and the discussion in Baraffe et al. 1998).
If the surface temperature of the star is reduced from $T_0$ to $T$
then to maintain a fixed luminosity the radius increases as $R = R_0
(T_0/T)^{1/2}$. Whilst this does not effect the bolometric magnitude,
it does affect $M_I$ and $M_K$, since the bolometric corrections,
$BC_I$ and $BC_K$, are functions of $T$ and $R$. Changes in $BC_I$ and
$BC_K$ with $T$ are found from the 158\,Myr isochrone. The second order
effect due to larger $R$ (and hence lower surface gravity) is estimated
by interpolating model isochrones for M-dwarfs from 15.8 to
158\,Myr. Bolometric corrections at $T$ are used to determine modified
isochrones and hence $M_I$ and $M_K$ for a star at the spot
temperature. A fraction $\lambda$, of this star is combined with a
fraction $(1-\lambda)$ of an unspotted star of similar radius to
determine the magnitude and colour of the spotted star as a function of
radius.

Two dashed histograms in Fig.~3 present results for representative
experiments. Calculated radii at the measured $M_I$ and $(I-K)_0$ are
binned to compare with the measurements. The first model shows the
result of a uniform 15 per cent temperature reduction in the models
(i.e. $\lambda=1$). This provides a reasonable match for radius vs
$M_I$, but fails to predict the correct stellar colours in Fig.~3b,
because the temperature dependence of $BC_K$ is very different to that
of $BC_I$.  Instead we simulate the radii of heavily spotted stars with
$\lambda=0.5$ and spots that are 30 per cent cooler than the unspotted
photosphere; at the upper end of spectroscopically deduced spot
coverages for active stars (e.g. O'Neal et al. 2004). The match between
predicted and measured radii is much improved in the radius vs $M_I$
plot {\it and} the radius vs $(I-K)_0$ plot. Even better agreement to the
lower luminosity bins can be achieved using $\lambda \simeq 0.75$ and
45 per cent spot temperature reductions. Such high filling factors are
suggested by investigations of Zeeman broadening due to magnetic fields
in rapidly rotating M-dwarfs (e.g Johns-Krull \& Valenti 2000; Reiners,
Basri \& Browning 2009) However, this level of precision is probably
over-interpretation at this stage. The remaining discrepancies could
also be due to uncertainties in the cluster metallicity or inaccuracies
in the unspotted model isochrone of radius versus $(I-K)_0$.

\section{Summary}

The results reported here show that the radii of young, single, rapidly
rotating M-dwarfs in NGC~2516 are significantly larger, at a given
luminosity, than predicted by evolutionary models and hence their
effective temperatures must be cooler than predicted. The
discrepancy becomes larger for stars less luminous than the predicted
transition to fully convective structures, reaching a radius anomaly 
of 50 per cent. These results are consistent, although more
extreme for the lowest mass stars in our sample, with radius anomalies
found for the components of older, tidally locked eclipsing binary
systems.  Evolutionary models correctly predict the radii of
magnetically in-active stars, suggesting that the increased radii in our
sample is caused by rapid rotation and/or magnetic activity. A simple,
two-temperature starspot model could explain the overall radius
increase, but only if spots cover $\geq 50$ per cent of the 
surface for the lowest luminosity objects in our sample.

Cool stars may be rapidly rotating and magnetically active as a
consequence of their youth or membership of tidally locked binary
systems. An altered relationship between mass, radius and effective
temperature for these stars calls for improvements in low-mass models
in order to interpret observations of young clusters, star forming
regions and close binary systems.

\section*{Acknowledgements}
Based on observations collected at the European Southern Observatory,
Paranal, Chile through observing programs 380.D-0479 and 266.D-5655. 
RJJ acknowledges receipt of an STFC studentship.

\nocite{Baraffe1998a}
\nocite{Baraffe2002a}
\nocite{Bagnulo2003a}
\nocite{Berger2006a}
\nocite{Carpenter2001a}   
\nocite{Chabrier2007a}
\nocite{Claret1995a}
\nocite{Cutri2003a}
\nocite{Dantona1997a}
\nocite{DAntona2000a}
\nocite{Demory2009a} 
\nocite{Hawley1999a}
\nocite{Horne1986a} 
\nocite{Irwin2007a}
\nocite{James2000a}
\nocite{Jeffries2001a}
\nocite{Jeffries1998a}
\nocite{Jeffries2007a}
\nocite{Leggett1992a}
\nocite{LopezMorales2007a}
\nocite{Lyra2006a}
\nocite{Morales2009a}
\nocite{Mullan2001a}
\nocite{ONeal2004a}
\nocite{Siess2000a}
\nocite{Terndrup2002a}
\nocite{Delfosse1998a}
\nocite{Reiners2009a}
\nocite{JohnsKrull2000a}
   
\bibliographystyle{mn2e} 
\bibliography{RJJbib}

\begin{thebibliography}{}

\bibitem[\protect\citeauthoryear{{Bagnulo, S. et al.}}{{Bagnulo, S. et
  al.}}{2003}]{Bagnulo2003a}
{Bagnulo, S. et al.} 2003, Messenger, 114, 10

\bibitem[\protect\citeauthoryear{{Baraffe}, {Chabrier}, {Allard} \&
  {Hauschildt}}{{Baraffe} et~al.}{1998}]{Baraffe1998a}
{Baraffe} I.,  {Chabrier} G.,  {Allard} F.,    {Hauschildt} P.~H.,  1998, \aap,
  337, 403

\bibitem[\protect\citeauthoryear{Baraffe, Chabrier, Allard \&
  Hauschildt}{Baraffe et~al.}{2002}]{Baraffe2002a}
Baraffe I.,  Chabrier G.,  Allard F.,    Hauschildt P.~H.,  2002, \aap, 382,
  563

\bibitem[\protect\citeauthoryear{{Berger, D.~H. et al.}}{{Berger, D.~H. et
  al.}}{2006}]{Berger2006a}
{Berger, D.~H. et al.} 2006, \apj, 644, 475

\bibitem[\protect\citeauthoryear{{Carpenter}}{{Carpenter}}{2001}]{Carpenter200%
1a}
{Carpenter} J.~M.,  2001, AJ, 121, 2851

\bibitem[\protect\citeauthoryear{{Chabrier}, {Gallardo} \&
  {Baraffe}}{{Chabrier} et~al.}{2007}]{Chabrier2007a}
{Chabrier} G.,  {Gallardo} J.,    {Baraffe} I.,  2007, \aap, 472, L17

\bibitem[\protect\citeauthoryear{{Claret}, {Diaz-Cordoves} \&
  {Gimenez}}{{Claret} et~al.}{1995}]{Claret1995a}
{Claret} A.,  {Diaz-Cordoves} J.,    {Gimenez} A.,  1995, \aaps, 114, 247

\bibitem[\protect\citeauthoryear{{Cutri, R. M. et al.}}{{Cutri, R. M. et
  al.}}{2003}]{Cutri2003a}
{Cutri, R. M. et al.} 2003, Technical report, Explanatory supplement to the
  2MASS All Sky data release.
http://www.ipac.caltech.edu/2mass/

\bibitem[\protect\citeauthoryear{{D'Antona} \& {Mazzitelli}}{{D'Antona} \&
  {Mazzitelli}}{1997}]{Dantona1997a}
{D'Antona} F.,  {Mazzitelli} I.,  1997, Memorie della Societa Astronomica
  Italiana, 68, 807

\bibitem[\protect\citeauthoryear{{D'Antona}, {Ventura} \&
  {Mazzitelli}}{{D'Antona} et~al.}{2000}]{DAntona2000a}
{D'Antona} F.,  {Ventura} P.,    {Mazzitelli} I.,  2000, \apjl, 543, L77

\bibitem[\protect\citeauthoryear{{Delfosse}, {Forveille}, {Perrier} \&
  {Mayor}}{{Delfosse} et~al.}{1998}]{Delfosse1998a}
{Delfosse} X.,  {Forveille} T.,  {Perrier} C.,    {Mayor} M.,  1998, \aap, 331,
  581

\bibitem[\protect\citeauthoryear{{Demory, B.~-O et al.}}{{Demory, B.~-O et
  al.}}{2009}]{Demory2009a}
{Demory, B.~-O et al.} 2009, arXiv:0906.0602

\bibitem[\protect\citeauthoryear{{Hawley}, {Tourtellot} \& {Reid}}{{Hawley}
  et~al.}{1999}]{Hawley1999a}
{Hawley} S.~L.,  {Tourtellot} J.~G.,    {Reid} I.~N.,  1999, \aj, 117, 1341

\bibitem[\protect\citeauthoryear{{Horne}}{{Horne}}{1986}]{Horne1986a}
{Horne} K.,  1986, \pasp, 98, 609

\bibitem[\protect\citeauthoryear{{Irwin}, {Hodgkin}, {Aigrain}, {Hebb},
  {Bouvier}, {Clarke}, {Moraux} \& {Bramich}}{{Irwin}
  et~al.}{2007}]{Irwin2007a}
{Irwin} J.,  {Hodgkin} S.,  {Aigrain} S.,  {Hebb} L.,  {Bouvier} J.,  {Clarke}
  C.,  {Moraux} E.,    {Bramich} D.~M.,  2007, \mnras, 377, 741

\bibitem[\protect\citeauthoryear{{James}, {Jardine}, {Jeffries}, {Randich},
  {Collier Cameron} \& {Ferreira}}{{James} et~al.}{2000}]{James2000a}
{James} D.~J.,  {Jardine} M.~M.,  {Jeffries} R.~D.,  {Randich} S.,  {Collier
  Cameron} A.,    {Ferreira} M.,  2000, \mnras, 318, 1217

\bibitem[\protect\citeauthoryear{{Jeffries}}{{Jeffries}}{2007}]{Jeffries2007a}
{Jeffries} R.~D.,  2007, MNRAS, 381, 1169

\bibitem[\protect\citeauthoryear{{Jeffries}, {James} \& {Thurston}}{{Jeffries}
  et~al.}{1998}]{Jeffries1998a}
{Jeffries} R.~D.,  {James} D.~J.,    {Thurston} M.~R.,  1998, \mnras, 300, 550

\bibitem[\protect\citeauthoryear{{Jeffries}, {Thurston} \& {Hambly}}{{Jeffries}
  et~al.}{2001}]{Jeffries2001a}
{Jeffries} R.~D.,  {Thurston} M.~R.,    {Hambly} N.~C.,  2001, \aap, 375, 863

\bibitem[\protect\citeauthoryear{{Johns-Krull} \& {Valenti}}{{Johns-Krull} \&
  {Valenti}}{2000}]{JohnsKrull2000a}
{Johns-Krull} C.~M.,  {Valenti} J.~A.,  2000, in {Pallavicini} R.,  {Micela}
  G.,   {Sciortino} S.,  eds, Stellar Clusters and Associations: Convection,
  Rotation, and Dynamos Vol.~198 of Astronomical Society of the Pacific
  Conference Series.
pp 371--+

\bibitem[\protect\citeauthoryear{{Leggett}}{{Leggett}}{1992}]{Leggett1992a}
{Leggett} S.~K.,  1992, \apjs, 82, 351

\bibitem[\protect\citeauthoryear{{L{\'o}pez-Morales}}{{L{\'o}pez-Morales}}{200%
7}]{LopezMorales2007a}
{L{\'o}pez-Morales} M.,  2007, \apj, 660, 732

\bibitem[\protect\citeauthoryear{{Lyra}, {Moitinho}, {van der Bliek} \&
  {Alves}}{{Lyra} et~al.}{2006}]{Lyra2006a}
{Lyra} W.,  {Moitinho} A.,  {van der Bliek} N.~S.,    {Alves} J.,  2006, \aap,
  453, 101

\bibitem[\protect\citeauthoryear{{Morales, J.~C. et al.}}{{Morales, J.~C. et
  al.}}{2009}]{Morales2009a}
{Morales, J.~C. et al.} 2009, \apj, 691, 1400

\bibitem[\protect\citeauthoryear{{Mullan} \& {MacDonald}}{{Mullan} \&
  {MacDonald}}{2001}]{Mullan2001a}
{Mullan} D.~J.,  {MacDonald} J.,  2001, \apj, 559, 353

\bibitem[\protect\citeauthoryear{{O'Neal}, {Neff}, {Saar} \& {Cuntz}}{{O'Neal}
  et~al.}{2004}]{ONeal2004a}
{O'Neal} D.,  {Neff} J.~E.,  {Saar} S.~H.,    {Cuntz} M.,  2004, \aj, 128, 1802

\bibitem[\protect\citeauthoryear{{Reiners}, {Basri} \& {Browning}}{{Reiners}
  et~al.}{2009}]{Reiners2009a}
{Reiners} A.,  {Basri} G.,    {Browning} M.,  2009, \apj, 692, 538

\bibitem[\protect\citeauthoryear{{Siess}, {Dufour} \& {Forestini}}{{Siess}
  et~al.}{2000}]{Siess2000a}
{Siess} L.,  {Dufour} E.,    {Forestini} M.,  2000, \aap, 358, 593

\bibitem[\protect\citeauthoryear{{Terndrup}, {Pinsonneault}, {Jeffries},
  {Ford}, {Stauffer} \& {Sills}}{{Terndrup} et~al.}{2002}]{Terndrup2002a}
{Terndrup} D.~M.,  {Pinsonneault} M.,  {Jeffries} R.~D.,  {Ford} A.,
  {Stauffer} J.~R.,    {Sills} A.,  2002, \apj, 576, 950

\end{thebibliography}


\bsp 

\label{lastpage}

\end{document}